# Conditional probability, three-slit experiments, and the Jordan algebra structure of quantum mechanics

Gerd Niestegge

Zillertalstrasse 39, 81373 Muenchen, Germany
gerd.niestegge@web.de

*Abstract.* Most quantum logics do not allow for a reasonable calculus of conditional probability. However, those ones which do so provide a very general and rich mathematical structure, including classical probabilities, quantum mechanics as well as Jordan algebras. This structure exhibits some similarities with Alfsen and Shultz's non-commutative spectral theory, but these two mathematical approaches are not identical. Barnum, Emerson and Ududec adapted the concept of higher-order interference, introduced by Sorkin in 1994, into a general probabilistic framework. Their adaption is used here to reveal a close link between the existence of the Jordan product and the non-existence of interference of third or higher order in those quantum logics which entail a reasonable calculus of conditional probability. The complete characterization of the Jordan algebraic structure requires the following three further postulates: a Hahn-Jordan decomposition property for the states, a polynomial functional calculus for the observables, and the positivity of the square of an observable. While classical probabilities are characterized by the absence of any kind of interference, the absence of interference of third (and higher) order thus characterizes a probability calculus which comes close to quantum mechanics, but still includes the exceptional Jordan algebras.

# 1   Introduction

The interference manifested in the two-slit experiments with small particles is one of the best known and most typical quantum phenomena. It is somewhat surprising therefore that quantum mechanics rules out third-order interference. This was discovered by Sorkin [1] considering measures on the "sets of histories" with experimental set-ups like the well-known two-slit experiments, but with three and more slits. He introduced the interference terms $I_2$ and $I_3$ and detected that, although the second-order interference is a typical quantum phenomenon ($I_2 \neq 0$), the third-order interference does not occur in quantum mechanics ($I_3 = 0$).

In the present paper, Sorkin's interference terms $I_2$ and $I_3$ are ported to the framework of quantum logics with unique conditional probabilities which was introduced by the author in [2] and [3]. In [3] it was shown that each such quantum logic can be embedded in an order-unit space where a specific type of positive projections then represent the probability conditionalization similar to the Lüders – von Neumann quantum measurement process.

In this general framework, the identity $I_3=0$ is not automatically given, and its role in a reconstruction of quantum mechanics from a few basic principles or in an axiomatic access to quantum mechanics based on a few interpretable postulates, are analysed in the paper. It is shown that the absence of third-order interference ($I_3=0$) has some important consequences. It entails the existence of a product in the order-unit space generated by the quantum logic, which can be used to characterize those quantum logics that can be embedded in the projection lattice in a Jordan algebra. Most of these Jordan algebras can be represented as operator algebras on a Hilbert space, and a reconstruction of quantum mechanics up to this point is thus achieved.

Besides the identity $I_3=0$, two further typical properties of quantum mechanics distinguishing it from more general theories are identified; these are a novel bound for quantum interference and a symmetry property of the conditional probabilities. This latter property was discovered by Alfsen and Shultz who used it as a postulate to derive the Jordan product for the quantum mechanical observables from it in their approach [4], and it was used in a similar way in [3], but a physical justification for it is hard to find. With the main result of the present paper, it can now be replaced by another postulate with a clearer physical meaning - namely the absence of third-order interference ($I_3=0$).

The next two sections summarize those parts of [2] and [3] which are relevant for the subsequent sections. In section 4, the second- and third-order interference terms ($I_2$ and $I_3$) are considered and ported to the quantum logics with unique conditional probabilities. The bound for quantum interference and the symmetry property of the quantum mechanical conditional probabilities are studied in sections 5 and 6. In section 7, a useful type of linear maps is introduced, which is used in section 8 to analyse the case $I_3=0$. A certain mathematical condition - the Jordan decomposition property - is outlined in section 9 and then used in section 10 to derive the product in the order-unit space from the identity $I_3=0$. Section 11 finally addresses the question under which further conditions the order-unit space becomes a Jordan algebra.

# 2   Quantum logics with unique conditional probabilities

A quantum logic is the mathematical model of a system of quantum events or propositions. Logical approaches use the name "proposition", while the name "event" is used in probability theory and will also be preferred in the present paper. The concrete quantum logic of standard quantum mechanics is the system of closed linear subspaces of a Hilbert space or, more generally, the projection lattice in a von Neumann algebra.

Usually, an abstract quantum logic is assumed to be an orthomodular partially ordered set and, very often, it is also assumed that it is lattice. For the purpose of the present paper, however, a more general and simpler mathematical structure without order relation is sufficient. Only an orthocomplementation, an orthogonality relation and a sum operation defined for orthogonal events are needed. The orthocomplementation represents the logical negation, orthogonality means mutual exclusivity, and the sum represents the logical and-operation in the case of mutual exclusivity. The precise axioms were presented in [2] and look as follows.

The quantum logic $E$ is a set with distinguished elements $0$ and $\mathbb{1}$, an orthogonality relation $\perp$ and a partial binary operation $+$ such that the following six axioms hold for $e,f,g \in E$:

(OS1)  *If $e \perp f$, then $f \perp e$; i.e., the relation $\perp$ is symmetric.*
(OS2)  *$e+f$ is defined for $e \perp f$, and then $e+f=f+e$; i.e., the sum operation is commutative.*
(OS3)  *If $g \perp e$, $g \perp f$, and $e \perp f$, then $g \perp e+f$, $f \perp g+e$ and $g+(e+f)=(g+e)+f$; i.e., the sum operation is associative.*
(OS4)  *$0 \perp e$ and $e+0=e$ for all $e \in E$.*
(OS5)  *For every $e \in E$, there exists a unique $e' \in E$ such that $e \perp e'$ and $e+e'=\mathbb{1}$.*
(OS6)  *There exists $d \in E$ such that $e \perp d$ and $e+d=f$ if and only if $e \perp f'$.*

Then $0'=\mathbb{1}$ and $e''=e$ for $e \in E$. Note that an orthomodular partially ordered set satisfies these axioms with the two definitions

(i)  *$e \perp f$ iff $f \leq e'$*
(ii) *The sum $e+f$ is the supremum of $e$ and $f$ for $e \perp f$.*

The supremum exists in this case due to the orthomodularity.

A state is a map $\mu: E \to [0,1]$ such that $\mu(\mathbb{1})=1$ and $\mu(e+f) = \mu(e) + \mu(f)$ for orthogonal pairs $e$ and $f$ in $E$. Then $\mu(0)=0$ and $\mu(e_1+...+e_k) = \mu(e_1)+...+\mu(e_k)$ for mutually orthogonal elements $e_1,...,e_k$ in $E$. Denote by $S$ the set of all states on $E$. With a state $\mu$ and $\mu(e)>0$ for an $e \in E$, another state $\nu$ is called a conditional probability of $\mu$ under $e$ if $\nu(f) = \mu(f)/\mu(e)$ holds for all $f \in E$ with $f \perp e'$. Furthermore, the following axioms were introduced in [2].

(UC1) *If $e,f \in E$ and $\mu(e)=\mu(f)$ for all $\mu \in S$, then $e=f$.*
(UC2) *If $e \in E$ and $\mu \in S$ with $\mu(e)>0$, there is one and only one conditional probability of $\mu$ under $e$.*

If these axioms are satisfied, $E$ is called a *UCP space* - named after the major feature of this mathematical structure which is the existence of the <u>u</u>nique <u>c</u>onditional <u>p</u>robability - and the elements in $E$ are called events. The unique conditional probability of $\mu$ under $e$ is denoted by $\mu_e$ and, in analogy with classical mathematical probability theory, $\mu(f|e)$ is often written instead of $\mu_e(f)$ with $f \in E$. The above two axioms imply that there is a state $\mu \in S$ with $\mu(e)=1$ for each event $e \neq 0$, that the difference $d$ in (OS6) becomes unique, and that $e \perp e$ iff $e \perp \mathbb{1}$ iff $e=0$ ($e \in E$).

Note that the following identity which will be used later holds for convex combinations of states $\mu,\nu \in S$ ($0<s<1$):

$$(s\mu+(1-s)\nu)_e = \frac{1}{s\mu(e)+(1-s)\nu(e)}(s\mu(e)\mu_e+(1-s)\nu(e)\nu_e). \tag{1}$$

A typical example of the above structure is the projection lattice $E$ in a von Neumann algebra $M$ without type $I_2$ part; $E = \{e \in M: e^*=e=e^2\}$. The conditional probabilities then have the shape

$$\mu_e(f) = \mu(f|e) = \frac{1}{\mu(e)} \hat{\mu}(efe) \tag{2}$$

with $e,f \in E$, $\mu \in S$ and $\mu(e)>0$. Note that $\hat{\mu}$ on $M$ is the unique positive linear extension of the state $\mu$ originally defined only on the projection lattice; this extension exists by Gleason's theorem [5] and its later enhancements to finitely additive states and arbitrary von Neumann algebras [6], [7], [8], [9]. The linear extension $\hat{\mu}$ does not exist if $M$ contains a type $I_2$ part.

For the proof of equation (2), suppose that the state $\nu$ on $E$ is a version of the conditional probability of the state $\mu$ under $e$ and use the identity $f=efe+efe'+e'fe+e'fe'$. From $\nu(e')=0$ and the Cauchy-Schwarz inequalty applied with the positive linear functional $\hat{\nu}$ it follows $0 = \hat{\nu}(efe') = \hat{\nu}(e'fe) = \hat{\nu}(e'fe')$ and thus $\nu(f) = \hat{\nu}(efe)$. By the spectral theorem, $efe$ can be approximated (in the norm topology) by linear combinations of elements in $\{d \in E: d \perp e'\} = \{d \in E: d \leq e\}$ for which $\nu$ coincides with $\mu/\mu(e)$. The continuity of $\hat{\nu}$ (due its positivity) then implies $\nu(f) = \hat{\nu}(efe) = \hat{\mu}(efe)/\mu(e)$. Therefore the conditional probability must have this shape and its uniqueness is proved. Its existence follows from $efe \geq 0$ and $efe=f$ for $f \leq e$, since then $\nu(f) := \hat{\mu}(efe)/\mu(e)$ indeed owns all the properties of the conditional probability.

Equation (2) reveals the link to the Lüders - von Neumann quantum measurement process. The transition from a state $\mu$ to the conditional probability $\mu_e$ is identical with the transition from the state prior to the measurement to the state after the measurement where $e$ represents the measurement result.

## 3  The embedding of the quantum logic in an order-unit space

A quantum logic with a sufficiently rich state space as postulated by (UC1) can be embedded in the unit interval of an order-unit space. In the present section, it will be shown that the existence and the uniqueness of the conditional probabilities postulated by (UC2) give rise to some important additional structure on this order-unit space, which was originally presented in [3].

A partially ordered real vector space $A$ is an order-unit space if $A$ contains an Archimedean order unit $\mathbb{1}$ [10], [11], [12]. The order unit $\mathbb{1}$ is positive and, for all $a \in A$, there is $t>0$ such that $-t\mathbb{1} \leq a \leq t\mathbb{1}$. An order unit $\mathbb{1}$ is called Archimedean if $na \leq \mathbb{1}$ for all $n \in \mathbb{N}$ implies $a \leq 0$. An order-unit space $A$ has a norm given by $\|a\| = \inf\{t>0: -t\mathbb{1} \leq a \leq t\mathbb{1}\}$. Each $x \in A$ can be written as $x=a-b$ with positive $a,b \in A$ (e.g., choose $a = \|x\|\mathbb{1}$ and $b = \|x\|\mathbb{1} - x$). A positive linear functional $\rho: A \to \mathbb{R}$ on an order-unit space $A$ is bounded with $\|\rho\|=\rho(\mathbb{1})$ and, vice versa, a bounded linear functional $\rho$ with $\|\rho\|=\rho(\mathbb{1})$ is positive.

The order-unit space $A$ considered in the following is the dual space of a base-norm space $V$ and, therefore, the unit ball of $A$ is compact in the weak-*-topology $\sigma(A,V)$. For $\rho \in V$ and $x \in A$ define $\hat{\rho}(x) := x(\rho)$; the map $\rho \to \hat{\rho}$ is the canonical embedding of $V$ in its second dual $V^{**}=A^*$. Then $\rho \in V$ is positive iff $\hat{\rho}$ is positive on $A$.

For any set $K$ in $A$, denote by $\overline{lin}\,K$ the $\sigma(A,V)$-closed linear hull of $K$ and by $\overline{conv}\,K$ the $\sigma(A,V)$-closed convex hull of $K$. For a convex set $K$, denote by $ext\,K$ the set of its extreme points which may be empty unless $K$ is compact. A projection is a linear map $U:A \to A$ with $U^2=U$ and, for $a \leq b$, define $[a,b] := \{x \in A: a \leq x \leq b\}$. Suppose that $E$ is a subset of $[0,\mathbb{1}]$ in $A$ such that

    (a) $\mathbb{1} \in E$,
    (b) $\mathbb{1}-e \in E$ if $e \in E$, and
    (c) $d+e+f \in E$ if $d,e,f,d+e,d+f,e+f \in E$.

Define $e':=\mathbb{1}-e$ and call $e,f\in E$ orthogonal if $e+f\in E$. Then $E$ satisfies the axioms (OS1),...,(OS6), and the states on $E$ and the state space $S$ can be considered as in section 2.

**Proposition 3.1:** *Suppose that $A$ is an order-unit space with order unit $\mathbb{1}$ and that $A$ is the dual of the base-norm space $V$. Moreover, suppose that $E$ is a subset of $[0,\mathbb{1}]$ satisfying the three above conditions* (a), (b), (c) *and that the following two conditions hold*:
(i) $A = \overline{lin}\ E$, *and for each $\mu\in S$ there is a $\sigma(A,V)$-continuous positive linear functional $\hat{\mu}$ on $A$ with $\hat{\mu}(e)=\mu(e)$ for $e\in E$.*
(ii) *For each $e\in E$ there is a $\sigma(A,V)$-continuous positive projection $U_e:A\to A$ such that $U_e\mathbb{1}=e$, $U_eA = \overline{lin}\{f\in E: f\leq e\}$ and $\hat{\mu} = \hat{\mu} U_e$ for $\mu\in S$ with $\mu(e)=1$.*
*Then $E$ is a UCP space. The conditional probabilities have the shape $\mu(f|e)=\hat{\mu}(U_ef)/\mu(e)$ for $e,f\in E$ and $\mu\in S$ with $\mu(e)>0$.*

*Proof.* For $e,f\in E$ with $e\neq f$ there is $\rho\in V_+$ with $\rho(e-f)\neq 0$. The restriction of $\rho/\rho(\mathbb{1})$ to $E$ then yields a state $\mu\in S$ with $\mu(e)\neq\mu(f)$. Therefore (UC1) holds.

Suppose $e\in E$ and $\mu\in S$ with $\mu(e)>0$. It is rather obvious that the map $g\to\hat{\mu}(U_eg)/\mu(e)$ on $E$ provides a conditional probability of $\mu$ under $e$. Now assume that $\nu$ is a further conditional probability of $\mu$ under $e$. Then $\nu(e)=1$ and thus $\hat{\nu}=\hat{\nu}U_e$. From $U_eg\in\overline{lin}\{f\in E: f\leq e\}$ it follows that $\nu(g) = \hat{\nu}(U_eg) = \hat{\mu}(U_eg)/\mu(e)$ for $g\in E$. Therefore, (UC2) holds as well. □

Note that the linear extension $\hat{\mu}$ in (i) is unique since $A = \overline{lin}\ E$. It shall now be seen that the situation of Proposition 3.1 is universal for the quantum logics with unique conditional probabilities; i.e., each UCP space has such a shape as described there.

**Theorem 3.2:** *Each UCP space $E$ is a subset of the interval $[0,\mathbb{1}]$ in some order-unit space $A$ with predual $V$ as described in Proposition* 3.1.

*Proof.* Define $V:=\{s\mu-t\nu: \mu,\nu\in S,\ 0\leq s,t\in\mathbb{R}\}$, which is a linear subspace of the orthogonally additive real-valued functions on $E$, and consider for $\rho\in V$ the norm

$$\|\rho\| := inf\{\ r\in\mathbb{R}\ : r\geq 0 \text{ and } \rho\in r\ conv(S\cup -S)\}.$$

Then $|\rho(e)| \leq \|\rho\|$ for every $e\in E$. Let $A$ be the dual space of the base-norm space $V$ and let $\hat{\mu}$ be the canonical embedding of $\mu\in V$ in $V^{**}=A^*$. If $\hat{\mu}(x) \geq 0$ for all $\mu\in S$, the element $x\in A$ is called positive and in this case write $x\geq 0$. Equipped with this partial ordering, $A$ becomes an order-unit space with the order unit $\mathbb{1}:=\pi(1)$, and the order-unit norm of an element $x\in A$ is $\|x\| = \sup\{\ |\hat{\mu}(x)|\ :\mu\in S\ \}$. With $e\in E$ define $\pi(e)$ in $A$ via $\pi(e)(\rho) := \rho(e)$ for $\rho\in V$. Then $0 \leq \|\pi(e)\| \leq \mathbb{1}$, and $\pi(e+f)=\pi(e)+\pi(f)$ for two orthogonal events $e$ and $f$ in $E$. Moreover, $A$ is the $\sigma(A,V)$-closed linear hull of $\pi(E)$.

Now the positive projection $U_e$ shall be defined for $e\in E$. Suppose $x\in A$ and $s\mu-t\nu \in V$ with $\mu,\nu\in S$ and $0\leq s,t\in\mathbb{R}$. Then define $(U_ex)(s\mu-t\nu) := s\mu(e)\hat{\mu}_e(x)-t\nu(e)\hat{\nu}_e(x)$. Here, $\hat{\mu}_e$ and $\hat{\nu}_e$ are the canonical embeddings of the conditional probabilities $\mu_e$ and $\nu_e$ in $A^*$; they do not exist in the cases $\mu(e)=0$ or $\nu(e)=0$ and then define $\mu(e)\hat{\mu}_e(x) := 0$ and $\nu(e)\hat{\nu}_e(x) :=0$, respectively. It still has to be shown that $U_e$ is well defined for $s\mu-t\nu = s'\mu'-t'\nu'$ with $\mu,\mu',\nu,\nu'\in S$ and $0\leq s,s',t,t'$. Then $s-t = (s\mu-t\nu)(\mathbb{1}) = (s'\mu'-t'\nu')(\mathbb{1}) = s'-t'$ and $s+t'=s'+t$. If $s+t'=0$, $s=s'=t=t'=0$ and $U_ex$ is well-defined. If $s+t'>0$, then either $s\mu(e) + t'\nu'(e) = s'\mu'(e) + t\nu(e) = 0$ and $s\mu(e) = t\nu(e) = s'\mu'(e) = t\nu(e) = 0$, or $(s\mu+t'\nu')/(s+t') = (s'\mu'+t\nu)/(s'+t) \in S$ and, calculating the conditional probability under $e$ for both sides of this identity by using (1), yields $s\mu(e)\mu_e+t'\nu'(e)\nu'_e$

- 5 -

$= s'\mu'(e)\mu'_e + t\nu(e)\nu_e$. In all cases, $U_e$ is well defined.

If $\mu(e)=1$ for $\mu \in S$, then $\mu=\mu_e$ and $\hat{\mu}(U_e x) = (U_e x)(\mu) = \hat{\mu}(x)$; i.e., $\hat{\mu} = \hat{\mu}U_e$. Thus, $(U_e U_e x)(\mu) = \mu(e)\hat{\mu}_e(U_e x) = \mu(e)\hat{\mu}_e(x) = (U_e x)(\mu)$ for all $\mu \in S$ and hence for all $\rho \in V$. Therefore $U_e U_e = U_e$, i.e., $U_e$ is a projection. Its positivity, $\sigma(A,V)$-continuity as well as $U_e 1 = \pi(e)$ and $U_e \pi(f) = \pi(f)$ for $f \in E$ with $f \leq e$ follow immediately from the definition.

Therefore $\overline{lin}\{\pi(f): f \in E, f \leq e\} \subseteq U_e A$. Assume $U_e x \notin \overline{lin}\{\pi(f): f \in E, f \leq e\}$ for some $x \in A$. By the Hahn-Banach theorem, there is $\rho \in V$ with $\hat{\rho}(U_e x) \neq 0$ and $\rho(f)=0$ for $f \in E$ with $f \leq e$. Suppose $\rho = s\mu - t\nu$ with $\mu,\nu \in S$ and $0 \leq s,t \in \mathbb{R}$. Then $s\mu(f)=t\nu(f)$ for $f \in E$ with $f \leq e$ and thus $s\mu(e)\mu_e(f) = t\nu(f)\nu_e(f)$. The uniqueness of the conditional probability implies $s\mu(e)\mu_e(f) = t\nu(f)\nu_e(f)$, i.e., $\hat{\rho}(U_e f) = 0$ for all $f \in E$ and hence $\hat{\rho}U_e = 0$ which contradicts $\hat{\rho}(U_e x) \neq 0$. This completes the proof of Theorem 3.2 after identifying $\pi(E)$ with $E$. □

**Lemma 3.3:** *If $e$ and $f$ are events in a UCP space $E$ with $e \leq f$, then $U_f e = e = U_f e$ and $U_e U_f = U_f U_e = U_e$. If $e, f \in E$ are orthogonal, then $U_e f = 0 = U_f e$, $U_e U_f = U_f U_e = 0$ and $U_{e'}U_{f'} = U_{(e+f)'} = U_{f'}U_{e'}$.*

*Proof.* Suppose $e \leq f$ and $\mu \in S$. Then $1=\mu_e(e) \leq \mu_e(f) \leq 1$ for the conditional probability $\mu_e$ implies $\mu_e(f)=1$ and hence $\hat{\mu}_e U_f = \hat{\mu}_e$. Therefore $\hat{\mu}U_e U_f = \mu(e)\hat{\mu}_e U_f = \mu(e)\hat{\mu}_e = \hat{\mu}U_e$ for all $\mu \in S$ and thus $U_e U_f = U_e$. The identity $U_f U_e = U_e$ immediately follows from (ii) in Proposition 3.1. Moreover $e = U_e 1 = U_e U_f 1 = U_e f$ and $e = U_e 1 = U_f U_e 1 = U_f e$.

Now assume that $e$ and $f$ are orthogonal. Then $e \leq f'$. Hence $e=U_e f' = U_e(1-f)=e-U_e f$, and $U_e f=0$. In the same way it follows that $U_f e=0$. Therefore $U_f$ vanishes on $U_e A = \overline{lin}\{d \in E: d \leq e\}$ and $U_f U_e=0$. The identity $U_e U_f=0$ follows in the same way.

Moreover, $0 \leq U_{e'}U_{f'}x \leq U_{e'}U_{f'}1 = U_{e'}f' = U_{e'}1 - U_{e'}f = e'-f = (e+f)'$ for $x \in [0,1]$. Therefore $\hat{\mu}U_{e'}U_{f'} = 0 = \hat{\mu}U_{(e+f)'}$ for $\mu \in S$ with $\mu((e+f)')=0$. Now consider $\mu \in S$ with $\mu((e+f)') > 0$ and define $\nu := \hat{\mu}U_{e'}U_{f'}/\mu((e+f)') \in S$. From $(e+f)' \leq e'$ and $(e+f)' \leq f'$, it follows that $U_{e'}U_{f'}(e+f)' = (e+f)'$ and $\nu((e+f)') = 1$. Hence $\nu = \hat{\nu}U_{(e+f)'}$. From $U_{e'}U_{(e+f)'} = U_{(e+f)'} = U_{f'}U_{(e+f)'}$ it follows that $\nu = \hat{\mu}U_{(e+f)'}/\mu((e+f)')$ and thus $\hat{\mu}U_{e'}U_{f'} = \hat{\mu}U_{(e+f)'}$. This identity now holds for all states $\mu$ and therefore $U_{e'}U_{f'}=U_{(e+f)'}$. In the same way it follows that $U_{f'}U_{e'}=U_{(e+f)'}$. □

The projections $U_e$ considered here are similar to, but not identical with the so-called P-projections considered by Alfsen and Shultz in their non-commutative spectral theory [13]. A P-projection $P$ has a quasicomplement $Q$ such that $Px=x$ iff $Qx=0$ (and $Qx=x$ iff $Px=0$) for $x \geq 0$. If $U_e x=x$, then $U_{e'}x=U_{e'}U_e x=0$ by Lemma 3.3, but $U_{e'}x=0$ does not imply $U_e x=x$.

An element $P(1)$ with a P-projection $P$ is called a projective unit by Alfsen and Shultz. In the case of spectral duality of a base-norm space $V$ and an order-unit space $A$, the system of projective units in $A$ is a UCP space if each state on the projective units has a linear extension to $A$ (as with the Gleason theorem, or in condition (i) of Proposition 3.1).

## 4 The interference terms $I_2$ and $I_3$

With two disjoint events $e_1$ and $e_2$, the classical conditional probabilities satisfy the rule $\mu(f|e_1+e_2)\mu(e_1+e_2) = \mu(f|e_1)\mu(e_1)+\mu(f|e_2)\mu(e_2)$. Only because violating this rule, quantum mechanics can correctly model the quantum interference phenomena observed in nature with small particles.

For instance, consider the two-slit experiment and let $e_1$ be the event that the particle passes through the first slit, $e_2$ the event that it passes through the second slit, and $f$ the event that it is registered in a detector located at a fixed position somewhere behind the screen with the two slits. Then $\mu(f|e_1)\mu(e_1)$ is the probability that the particle is registered in the detector when the first slit is open and the second one is closed, $\mu(f|e_2)\mu(e_2)$ is the probability that the particle is registered in the detector when the second slit is open and the first one is closed, and $\mu(f|e_1+e_2)\mu(e_1+e_2)$ is the probability that the particle is registered in the detector when both slits are open. If the above rule were valid, it would rule out the interference patterns observed in the quantum physical experiments and correctly modelled by quantum theory. Therefore, a first interference term is defined by

$$I_2^{\mu,f}(e_1,e_2) := \mu(f|e_1+e_2)\mu(e_1+e_2) - \mu(f|e_1)\mu(e_1) - \mu(f|e_2)\mu(e_2),$$

where $\mu$ is a state, $f$ any event, and $e_1,e_2$ an orthogonal pair of events. While $I_2^{\mu,f}(e_1,e_2) = 0$ in the classical case, it is typical of quantum mechanics that $I_2^{\mu,f}(e_1,e_2) \neq 0$.

With three orthogonal events $e_1,e_2,e_3$, a next interference term can be defined in the following way:

$$I_3^{\mu,f}(e_1,e_2,e_3) :=$$

$$\mu(f|e_1+e_2+e_3)\mu(e_1+e_2+e_3)$$

$$- \mu(f|e_1+e_2)\mu(e_1+e_2) - \mu(f|e_1+e_3)\mu(e_1+e_3) - \mu(f|e_2+e_3)\mu(e_2+e_3)$$

$$+ \mu(f|e_1)\mu(e_1) + \mu(f|e_2)\mu(e_2) + \mu(f|e_3)\mu(e_3)$$

Similar to the two-slit experiment, now consider an experiment where the screen has three instead of two slits. Let $e_k$ be the event that the particle passes through the $k^{th}$ slit ($k=1,2,3$) and $f$ again the event that it is registered in a detector located somewhere behind the screen with the slits. Then $\mu(f|e_k)\mu(e_k)$ is the probability that the particle is registered in the detector when the $k^{th}$ slit is open and the other two ones are closed, $\mu(f|e_i+e_j)\mu(e_i+e_j)$ is the probability that the particle is registered in the detector when the $i^{th}$ slit and the $j^{th}$ are open and the remaining third slit is closed, and $\mu(f|e_1+e_2+e_3)\mu(e_1+e_2+e_3)$ is the probability that the particle is registered in the detector when all three slits are open. The interference term $I_3^{\mu,f}(e_1,e_2,e_3)$ is the sum of these probabilities with negative signs in the cases with two slits open and positive signs in the cases with one or three slits open. The sum would become zero if these probabilities were additive in $e_1,e_2,e_3$ as they are in the classical case. However, this is not the only case; $I_3^{\mu,f}(e_1,e_2,e_3)=0$ means that the detection probability with three open slits is a simple linear combination of the detection probabilities in the three cases with two open slits and the three cases with one single open slit. If the situation with three slits involves some new interference, the detection probability should not be such a linear combination and $I_3^{\mu,f}(e_1,e_2,e_3)$ should not become zero. The fascinating question now arises whether $I_3^{\mu,f}(e_1,e_2,e_3) = 0$ or not.

Considering experiments with three and more slits, this third-order interference term and a whole sequence of further higher-order interference terms were introduced by Sorkin [1], but in another form. He used probability measures on "sets of histories". When porting his third-order interference term to conditional probabilities, it gets the above shape. The same shape is used by C. Ududec, H. Barnum and J. Emerson [14] who adapted Sorkin's third-order interference term into an operational probabilistic framework. The further higher-order interference terms will not be considered in the present paper. It may be interesting to note that the absence of the third-order interference implies the absence of the interference of all orders higher than three.

Using the identity $\mu(f|e)=\hat{\mu}(U_e f)/\mu(e)$, the validity of $I_3^{\mu,f}(e_1,e_2,e_3) = 0$ for all states $\mu$ is immediately equivalent to the identity

$$U_{e_1+e_2+e_3}f - U_{e_1+e_2}f - U_{e_2+e_3}f - U_{e_1+e_3}f + U_{e_1}f + U_{e_2}f + U_{e_3}f = 0.$$

If this shall hold also for all events $f$, this means

$$I_3(e_1,e_2,e_3) := U_{e_1+e_2+e_3} - U_{e_1+e_2} - U_{e_2+e_3} - U_{e_1+e_3} + U_{e_1} + U_{e_2} + U_{e_3} = 0. \quad (3)$$

The term $I_3(e_1,e_2,e_3)$ does not any more depend on a state or the event $f$, but only on the orthogonal event triple $e_1,e_2,e_3$. Note that $I_3(e_1,e_2,e_3)$ is a linear map on the order-unit space $A$ while $I_3^{\mu,f}(e_1,e_2,e_3) = \hat{\mu}(I_3(e_1,e_2,e_3)f)$ is a real number.

This interference term shall now be studied in a von Neumann algebra where the conditional probability has the shape $\mu(f|e)=\hat{\mu}(efe)/\mu(e)$ for projections $e,f$ and a state $\mu$ with $\mu(e)>0$, and hence $U_e f = efe$. Then

$$U_{e_1+e_2+e_3}f - U_{e_1+e_2}f - U_{e_2+e_3}f - U_{e_1+e_3}f + U_{e_1}f + U_{e_2}f + U_{e_3}f =$$

$$(e_1+e_2+e_3)f(e_1+e_2+e_3)$$

$$-(e_1+e_2)f(e_1+e_2) - (e_2+e_3)f(e_2+e_3) - (e_1+e_3)f(e_1+e_3)$$

$$+e_1 f e_1 + e_2 f e_2 + e_3 f e_3 = 0$$

Therefore, in a von Neumann algebra and in standard quantum mechanics, the identity $I_3^{\mu,f}(e_1,e_2,e_3) = 0$ always holds, which was already seen by Sorkin. This becomes a first interesting property of quantum mechanics distinguishing it from the general quantum logics with unique conditional probabilities (UCP spaces). It is quite surprising that quantum mechanics has this property because there is no obvious reason why $I_3^{\mu,f}(e_1,e_2,e_3)$ should vanish while $I_2^{\mu,f}(e_1,e_2)$ does not. Likewise surprising are the bounds which quantum mechanics imposes on $I_2^{\mu,f}(e_1,e_2)$ and which will be presented in the next section.

## 5  A bound for quantum interference

Suppose that $E$ is a UCP space. By Theorem 3.2, it can be embedded in an order-unit space $A$ such that Proposition 3.1 holds. For each event $e \in E$ define a linear map $S_e$ on $A$ by

$$S_e x := 2 U_e x + 2 U_{e'} x - x \quad (x \in A).$$

Then $S_e \mathbb{1} = \mathbb{1}$. Furthermore, $S_e S_{e'} x = S_{e'} S_e x = x$ for $x \in A$; i.e., $S_{e'}$ is the inverse of the linear map $S_e$ and $S_e$ is a linear isomorphism. If it were positive, it would be an automorphism of the order-unit space $A$, but this is not true in general. It shall now be studied what the positivity of the maps $S_e$ would mean for the conditional probabilities.

Suppose $e, f \in E$. Then $0 \leq S_e f$ means $f \leq 2U_e f + 2U_{e'} f$, and this is equivalent to the following inequality for the conditional probabilities

$$\mu(f) \leq 2\mu(f|e)\mu(e) + 2\mu(f|e')\mu(e') \tag{4}$$

or, equivalently,

$$I_2^{\mu,f}(e,e') \leq \mu(f|e)\mu(e) + \mu(f|e')\mu(e') \tag{5}$$

holding for all states $\mu$. Exchanging $f$ by $f'$ yields from $0 \leq S_e f'$

$$\begin{aligned} 1 - \mu(f) &= \mu(f') \leq 2\mu(f'|e)\mu(e) + 2\mu(f'|e')\mu(e') \\ &= 2(1-\mu(f|e))\mu(e) + 2(1-\mu(f|e'))\mu(e') \\ &= 2 - 2\mu(f|e)\mu(e) - 2\mu(f|e')\mu(e') \end{aligned}$$

and thus a second inequality

$$\mu(f) \geq 2\mu(f|e)\mu(e) + 2\mu(f|e')\mu(e') - 1 \tag{6}$$

or, equivalently,

$$I_2^{\mu,f}(e,e') \geq \mu(f|e)\mu(e) + \mu(f|e')\mu(e') - 1. \tag{7}$$

The above inequalities (5) and (7) introduce an upper bound and a lower bound for the interference term $I_2^{\mu,f}(e,e')$. How these bounds limit the interference, is shown in Figure 5.1 using the inequalities (4) and (6) and not this specific interference term. The dashed diagonal line represents the classical case without interference, while the inequalities (4) and (6) allow the whole corridor between the two continuous lines and forbid the area outside this corridor.

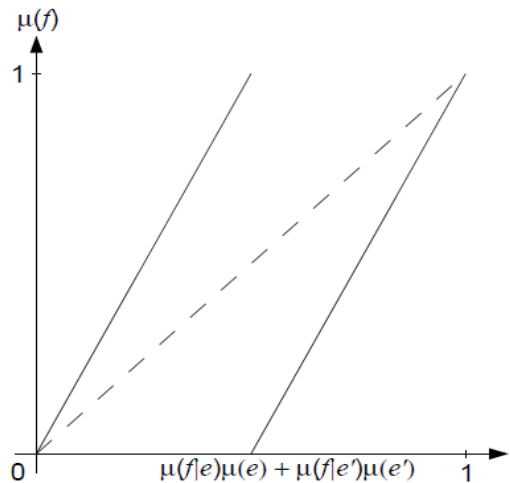

*Figure 5.1*

In a von Neumann algebra, $U_e f = efe$ for projections $e, f$ and hence $S_e x = 2exe + 2e'xe' - x = (e-e')x(e-e')$ with $x$ in the von Neumann algebra. Therefore the maps $S_e$ are positive and the above inequalities for the conditional probabilities ((4) and (6) - illustrated in Figure 5.1) and the resulting bounds for the interference term $I_2^{\mu,f}(e,e')$ (inequalities (5) and (7)) hold in this case. This is a second interesting property of quantum mechanics distinguishing it from other more general theories.

# 6 A symmetry property of the quantum mechanical conditional probabilities

Alfsen and Shultz introduced the following symmetry condition for the conditional probabilities in [4] and used it to derive a Jordan algebra structure from their non-commutative spectral theory:

(A1) $\quad\quad\quad\quad \mu(f'|e)\mu(e) + \mu(f|e')\mu(e') = \mu(e'|f)\mu(f) + \mu(e|f')\mu(f')$

This condition arose mathematically and a physical meaning is not immediately at hand. Alfsen and Shultz's interpretation was: „*The probability of the exclusive disjunction of two system properties is independent of the order of the measurements of the two system properties.*" However, the exclusive disjunction is not an event or proposition.

The first summand on the left-hand side $\mu(f'|e)\mu(e)$ is the probability that a first measurement testing $e$ versus $e'$ provides the result $e$ and that a second successive measurement testing $f$ versus $f'$ then provides the result $f'$ (= „not $f$"). The second summand on the left-hand side and the two summands on the right-hand side can be interpreted in the same way with exchanged roles of $e,e',f,f'$.

Figure 6.1 displays a measuring arrangement consisting of two successive measurements M1 and M2; the first one tests $e$ versus $e'$ and the second one $f$ versus $f'$. After the second one, the particle is let pass only in the two cases if the result of the first measurement is $e$ and the result of the second one is $f'$ or if the result of the first one is $e'$ and the result of the second one is $f$. In the other two cases, if the result of the first measurement is $e$ and the result of the second one is $f$ or if the result of the first one is $e'$ and the result of the second one is $f'$, the particle is absorbed after the second measurement. Figure 6.2 displays the same measuring arrangement as Figure 6.1, but with exchanged roles of $e$ and $f$.

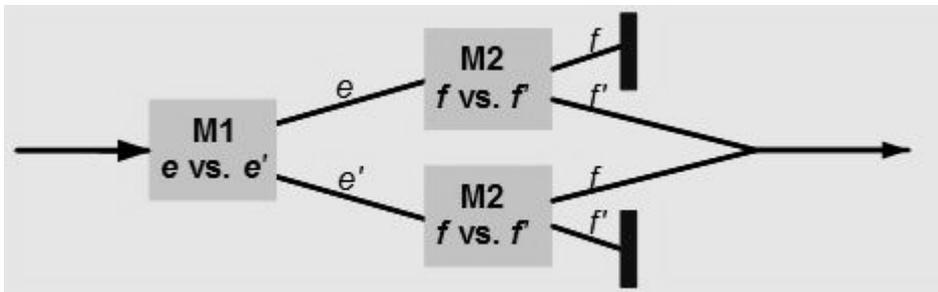

*Figure 6.1*

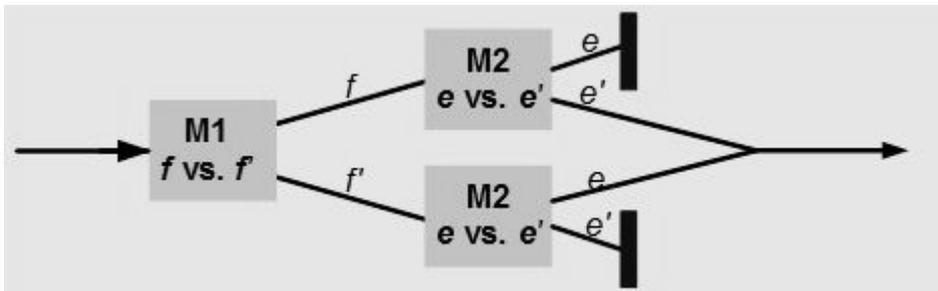

*Figure 6.2*

Condition (A1) now means that the probability that a particle passes through the measuring arrangement shown in Figure 6.1 is identical with the probability that a particle passes through the measuring arrangement shown in Figure 6.2.

The symmetry condition for the conditional probabilities (A1) also plays a certain role in the study of different compatibility/comeasurability levels in [15].

Condition (A1) shall now be rewritten using the interference term $I_2^{\mu,f}(e_1,e_2)$ for two orthogonal events $e_1$ and $e_2$. To remove the dependence on $f$ and the state $\mu$, first define in analogy to equation (3)

$$I_2(e_1,e_2) := U_{e_1+e_2} - U_{e_1} - U_{e_2} \tag{8}$$

which is a linear operator on the order-unit space $A$ generated by an UCP space $E$. Note that $I_2^{\mu,f}(e_1,e_2) = \hat{\mu}(I_2(e_1,e_2)f)$, recalling the identity $\mu(f|e)\mu(e) = \hat{\mu}(U_e f)$ for states $\mu$ and events $e$ and $f$. The validity of (A1) for all states $\mu$ is immediately equivalent to the identity

$$U_e f' + U_{e'} f = U_f e' + U_{f'} e . \tag{9}$$

It implies
$$\begin{aligned}
0 &= U_e f' + U_{e'} f - U_f e' - U_{f'} e \\
&= e - U_e f + U_{e'} f - f + U_f e - U_{f'} e \\
&= I_2(f,f')e + 2U_f e - I_2(e,e')f - 2U_e f
\end{aligned}$$

and hence

$$I_2(e,e')f - I_2(f,f')e = 2U_f e - 2U_e f . \tag{10}$$

Reconsidering the von Neumann algebras where $U_e f = efe$, equation (9) becomes $e(1-f)e + (1-e)f(1-e) = f(1-e)f + (1-f)e(1-f)$. Both sides of this equation are identical to $e+f-ef-fe$. Therefore (A1) holds for all states and all events in a von Neumann algebra and in the standard model of quantum mechanics. This is a third interesting property of quantum mechanics distinguishing it from the general quantum logics with unique conditional probabilities (UCP spaces). However, it is a mathematical property without a clear physical reason behind it and the absence of third-order interference ($I_3=0$) is the more interesting property from the physical point of view.

## 7  The linear maps $T_e$

In addition to the $U_e$, a further useful type of linear maps $T_e$ on the order-unit space $A$ generated by a UCP space $E$ shall now be defined for $e \in E$:

$$T_e x := \frac{1}{2}(x + U_e x - U_{e'} x), \quad x \in A. \tag{11}$$

In a von Neumann algebra, this becomes $T_e x = (ex+xe)/2$, which is the Jordan product of $e$ and $x$. This is a first reason why some relevance is expected from the maps $T_e$ on the order-unit space $A$.

A second reason is that (A1) holds for all states $\mu$ if and only if $T_e f = T_f e$. This follows from equation (9). Thus, using the linear maps $T_e$, the symmetry condition (A1) for the conditional probabilities is transformed to the very simple equation $T_e f = T_f e$. Actually, this is what brought Alfsen and Shultz to the discovery of (A1).

Some further important characteristics of these linear maps shall now be collected. Suppose $e \in E$ and $x \in A$. Using the above definition of $T_e$ immediately yields $T_e x + T_{e'} x = x$. Moreover,

$$T_e x = \frac{1}{2}(x - U_e x - U_{e'} x) + U_e x + 0 U_{e'} x$$

is the spectral decomposition of $T_e$ and hence $T_e$ has the three eigenvalues 0, ½, 1. Furthermore

$$T_e U_e x = \frac{1}{2}(U_e x + U_e^2 x - 0) = U_e x$$

and

$$T_e U_{e'} x = \frac{1}{2}(U_{e'} x + 0 - U_{e'}^2 x) = 0.$$

Therefore

$$T_e^2 x = \frac{1}{2}(T_e x + T_e U_e x - T_e U_{e'} x) = \frac{1}{2}(T_e x + U_e x)$$

and thus

$$U_e = 2 T_e^2 - T_e. \qquad (12)$$

Originally, the $T_e$ were derived from the $U_e$, and equation (12) means that the $U_e$ can be reconstructed from the $T_e$.

The norm of the operator $T_e$ shall now be calculated. From $U_e[-1,1] \subseteq [-e,e]$ and $U_{e'}[-1,1] \subseteq [-e',e']$ it follows that

$$(U_e - U_{e'})[-1,1] \subseteq [-1,1] \qquad (13)$$

and therefore $\|U_e - U_{e'}\| \leq 1$. Hence $\|T_e\| \leq 1$. Since $T_e e = e$ and $\|e\| = 1$ for $e \neq 0$, it follows that $\|T_e\| = 1$ unless $e = 0$ and $T_e = 0$.

**Lemma 7.1:** *If two events e and f in an UCP space E are orthogonal, then the linear maps $T_e$ and $T_f$ on the order-unit space A generated by E commute: $T_e T_f = T_f T_e$.*

*Proof.* By Lemma 3.3 the four projections $U_e$, $U_{e'}$, $U_f$, $U_{f'}$ commute pairwise. The linear maps $T_e$ and $T_f$ then commute by equation (11). □

An important link between the linear maps $T_e$ ($e \in E$) and Sorkin's interference term $I_3$ will be considered in the next sections.

## 8  Quantum logics with $I_3 = 0$

The interference term $I_2^{\mu,f}(e_1, e_2)$ vanishes for all states μ and all events $f$ if and only if $I_2(e_1, e_2) = U_{e_1+e_2} - U_{e_1} + U_{e_2} = 0$. The general validity of this identity for all orthogonal event pairs $e_1, e_2$ means that the map $e \to U_e$ is orthogonally additive in $e$. It will later be seen in Proposition 8.2 that the general validity of $I_3(e_1, e_2, e_3) = 0$ for all orthogonal event triples $e_1, e_2, e_3$ means that the map $e \to T_e$ is orthogonally additive in $e$. This will follow from the next lemma.

**Lemma 8.1:** *Suppose that E is a UCP space and that A is the order-unit space generated by E.*
(i) *If $e_1, e_2, e_3$ are three orthogonal events in E, then*

$$I_3(e_1, e_2, e_3) = U_{e_1+e_2+e_3} I_3((e_2+e_3)', e_2, e_3).$$

(ii) *If e and f are two orthogonal events in E, then*

$$T_e + T_f - T_{e+f} = \frac{1}{2} I_3(e, f, (e+f)').$$

*Proof.* (i) $\quad I_3((e_2+e_3)', e_2, e_3) = U_1 - U_{e_3'} - U_{e_2'} - U_{e_2+e_3} + U_{e_1} + U_{e_2} + U_{e_3}$

and then by Lemma 3.3

$$U_{e_1+e_2+e_3} I_3((e_2+e_3)', e_2, e_3) = U_{e_1+e_2+e_3} - U_{e_1+e_2} - U_{e_1+e_3} - U_{e_1+e_3} + U_{e_1} + U_{e_2} + U_{e_3}$$

$$= I_3(e_1, e_2, e_3).$$

(ii) With $e_1 := e$, $e_2 := f$ and $e_3 := (e+f)'$, it follows for $x \in A$:

$$T_{e+f} x = \frac{1}{2}(x + U_{e_1+e_2} x - U_{e_3} x)$$

$$T_e x = \frac{1}{2}(x + U_{e_1} x - U_{e_2+e_3} x)$$

$$T_f x = \frac{1}{2}(x + U_{e_2} x - U_{e_1+e_3} x)$$

and thus

$$T_e x + T_f x - T_{e+f} x = \frac{1}{2}(x + U_{e_1} x - U_{e_2+e_3} x) + \frac{1}{2}(x + U_{e_2} x - U_{e_1+e_3} x) - \frac{1}{2}(x + U_{e_1+e_2} x - U_{e_3} x)$$

$$= \frac{1}{2}(x - U_{e_1+e_2} x - U_{e_2+e_3} x - U_{e_1+e_3} x + U_{e_1} x + U_{e_2} x + U_{e_3} x)$$

$$= \frac{1}{2} I_3(e_1, e_2, e_3) x.$$

□

**Proposition 8.2:** *Suppose that E is a UCP space. Then the following three conditions are equivalent:*

(i) $\quad I_3(e_1, e_2, e_3) = 0 \quad$ *for all orthogonal events $e_1$, $e_2$, $e_3$ in E.*
(ii) $\quad I_3(e_1, e_2, e_3) = 0 \quad$ *for all orthogonal events $e_1$, $e_2$, $e_3$ in E with $e_1 + e_2 + e_3 = 1$.*
(iii) $\quad T_{e+f} = T_e + T_f \quad$ *for all orthogonal events e and f in E.*

*Proof.* The implication (i) $\Rightarrow$ (ii) is obvious, the implication (ii) $\Rightarrow$ (iii) follows from Lemma 8.1 (ii) and the implication (iii) $\Rightarrow$ (i) from Lemma 8.1 (i) and (ii) both. □

It now becomes clear that the symmetry condition (A1) for the conditional probabilities implies the absence of third-order interference, i.e., that $I_3(e_1,e_2,e_3)=0$ for all orthogonal events $e_1$, $e_2$, $e_3$. Recall that (A1) implies the identity $T_e f = T_f e$ for all events $e$ and $f$. Therefore, $T_e$ is orthogonally additive in $e$ and $I_3(e_1,e_2,e_3)=0$ follows from Proposition 8.2.

## 9  *Jordan decomposition*

In this section, the orthogonal additivity of $T_e$ in $e$ (by Lemma 7.2) will be used to define $T_x$ for all $x$ in the order-unit space $A$. The $T_e$ are not positive and this involves some difficulties, which can be overcome when the real-valued bounded orthogonally additive functions on the UCP space $E$ satisfy the so-called Jordan decomposition property or the stronger Hahn-Jordan decomposition property. These decomposition properties are named after the French mathematician Camille Jordan (1838 – 1922) who originally introduced the first one for functions of bounded variation and signed measures. The Jordan algebras are named after another person; this is the German physicist Pascual Jordan (1902 – 1980).

Consider functions $\rho\colon E \to \mathbb{R}$ which are orthogonally additive (i.e., $\rho(e+f)=\rho(e)+\rho(f)$ for orthogonal elements e and $f$ in E) and bounded (i.e., $sup\{|\rho(e)| : e\in E\} < \infty$). Let $R$ denote the set of all these functions $\rho$ on $E$. Then $R$ comprises the state space $S$. The UCP space $E$ is said to have the Jordan decomposition property if each $\rho\in R$ can be written in the form $\rho=s\mu-t\nu$ with two states $\mu$ and $\nu$ in $S$ and non-negative real number $s$ and $t$. It has the ε-Hahn-Jordan decomposition property if the following stronger condition holds: For each $\rho\in R$ and every ε>0 there are two states $\mu$ and $\nu$ in $S$, non-negative real number $s$ and $t$ and an event $e$ in $E$ such that $\rho=s\mu-t\nu$ and $\mu(e)<\varepsilon$ as well as $\nu(e')<\varepsilon$. This ε-Hahn-Jordan decomposition property was studied in the framework of quantum logics by Cook [16] and Rüttimann [17]. The usual Hahn-Jordan decomposition property for signed measures is even stronger requiring that $\mu(e)=0=\nu(e')$.

Note that the projection lattices of von Neumann algebras have the Jordan decomposition property and the ε-Hahn-Jordan decomposition property, but this is not obvious. It is well-known that theses types of decomposition are possible for the bounded linear functionals on the algebra, but they are needed for the orthogonally additive real functions on the projection lattice. Bunce and Maitland Wright [18] showed that each such function on the projection lattice has a bounded linear extension to the whole algebra (this is the last step of the solution of the Mackey-Gleason problem which had been open for a long time) and then the decomposition of this extension provides the desired decomposition by considering the restrictions of the linear functionals to $E$.

**Lemma 9.1:** *Suppose that E is a UCP space with the Jordan decomposition property and* $I_3(e_1,e_2,e_3)=0$ *for all orthogonal events $e_1$, $e_2$, $e_3$ in E. Then, for each x in the order-unit space A generated by E, the map $e\to T_e x$ from E to A has a unique $\sigma(A,V)$-continuous linear extension $y\to T_y x$ on A.*

*Proof.* Consider $V$ as defined in the proof of Theorem 3.2. In the general case, only the inclusion $V\subseteq R$ holds, but the Jordan decomposition property ensures that $V=R$.

For each $\rho$ in $V$ and each $x$ in $A$, define a function $\rho^x$ on $E$ by $\rho^x(e):=\hat{\rho}(T_e x)$. By Proposition 8.2, $\rho^x$ is orthogonally additive in $e$. Moreover, $|\rho^x(e)|\leq\|\rho\|\,\|T_e\|\,\|x\|\leq\|\rho\|\,\|x\|$ for $e\in E$. Thus $\rho^x$ is bounded and lies in $R=V$. Let $\hat{\rho}^x$ be its canonical embedding in $V^{**}=A^*$ and, with $y\in A$, consider the real-valued bounded linear map $\rho \to \hat{\rho}^x(y)$ on $V$. It defines an element $T_y x$ in $V^*=A$ such that the map $y \to T_y x$ is linear as well as $\sigma(A,V)$-continuous on $A$ and coincides with the original $T_e x$ for $y=e\in E$. □

## 10  The product on the order-unit space

Under the assumptions of the last lemma, a product can now be defined on the order-unit space $A$ by $y \square x := T_y x$. It is linear in $x$ as well as in $y$, but there is certain asymmetry concerning its $\sigma(A,V)$-continuity. The product $y \square x$ is $\sigma(A,V)$-continuous in $y \in A$ with $x \in A$ fixed, and it is $\sigma(A,V)$-continuous in $x \in A$ with $y = e \in E$ fixed, but generally not with other $y \in A$. Moreover, $1 \square x = x$ and $y \square 1 = y$, $e \square e = e$ and $\|e \square x\| \leq \|x\|$ for the elements $x$ and $y$ in $A$ and the events $e$ in $E$. However, the inequality $\|y \square x\| \leq \|y\| \|x\|$ is not yet available; this requires the $\varepsilon$-Hahn-Jordan decomposition property and will follow from the next lemma.

**Lemma 10.1:** *Suppose that $E$ is a UCP space with the $\varepsilon$-Hahn-Jordan decomposition property. Then $[-1,1]$ is identical with the $\sigma(A,V)$-closed convex hull of the set $\{e-e': e \in E\}$ in $A$, and $[0,1]$ coincides with the $\sigma(A,V)$-closed convex hull of $E$. Moreover, the extreme points of $[0,1]$ lie in the $\sigma(A,V)$-closure of $E$.*

*Proof.* The inclusion $\overline{conv}\{e-e': e \in E\} \subseteq [-1,1]$ is obvious, and it shall now be shown that $[-1,1] \subseteq \overline{conv}\{e-e': e \in E\}$. Assume that an $x$ exists in the interval $[-1,1]$ which does not lie in the set $\overline{conv}\{e-e': e \in E\}$. By the Hahn-Banach theorem, there is $\rho \in V$ with $\rho(x) > sup\{\rho(e-e'): e \in E\}$. From $0 = (e-e')/2 + (e'-e)/2$ for any $e \in E$, it follows that $0 \in \overline{conv}\{e-e': e \in E\}$. Hence $\rho(x) > 0$ and $\rho \neq 0$.

Suppose $\varepsilon > 0$. Due to the $\varepsilon$-Hahn-Jordan decomposition property, there are two states $\mu$ and $\nu$ in $S$, non-negative real number $s$ and $t$ and an event $f$ in $E$ such that $\rho = s\mu - t\nu$ and $\mu(f) < \varepsilon$ as well as $\nu(f') < \varepsilon$. Then

$$\rho(f'-f) = s\mu(f') - s\mu(f) - t\nu(f') + t\nu(f)$$

$$= s - 2s\mu(f) + t - 2t\nu(f')$$

$$\geq s+t - 2\varepsilon(s+t) = (1-2\varepsilon)(s+t).$$

Recall from the proof of Theorem 3.2 that $\|\rho\| = inf\{\ r \in \mathbb{R}\ : r \geq 0 \text{ and } \rho \in r\, conv(S \cup -S)\}$. The case $s=t=0$ cannot occur since this would imply $\rho=0$. Therefore write

$$\rho = (s+t)\left(\frac{s}{s+t}\mu - \frac{t}{s+t}\nu\right)$$

to see that $\|\rho\| \leq s+t$.

Finally $\rho(x) \leq \|\rho\| \|x\| \leq \|\rho\| \leq s+t \leq \rho(f'-f)/(1-2\varepsilon)$. With $\varepsilon$ small enough such that $(1-2\varepsilon)\rho(x) > sup\{\rho(e-e'): e \in E\}$, an event $f$ in $E$ is found with $\rho(f'-f) \geq (1-2\varepsilon)\rho(x) > sup\{\rho(e-e'): e \in E\}$, which is the desired contradiction with $e := f'$.

Thus $[-1,1] = \overline{conv}\{e-e': e \in E\}$. The map $x \to (x+1)/2$ is a $\sigma(A,V)$-continuous affine isomorphism mapping $[-1,1]$ to $[0,1]$ and $\{e-e': e \in E\}$ to $E$; therefore $[0,1] = \overline{conv}\ E$. The Krein-Milman theorem then yields that the extreme points of $[0,1]$ lie in the $\sigma(A,V)$-closure of $E$. □

**Lemma 10.2:** *Suppose that $E$ is a UCP space with the $\varepsilon$-Hahn-Jordan decomposition property and $I_3(e_1, e_2, e_3) = 0$ for all orthogonal events $e_1, e_2, e_3$ in $E$. Then $\|y \square x\| \leq \|y\| \|x\|$ for all elements $x, y$ in $A$.*

*Proof.* Suppose $e \in E$ and $x \in A$ with $\|x\| \leq 1$. I.e., $x \in [-\mathbb{1}, \mathbb{1}]$. Then $\|(e-e')\square x\| = \|e\square x - e'\square x\| = \|T_e x - T_{e'} x\|$. Moreover, $T_e x - T_{e'} x = U_e x - U_{e'} x$ by equation (11) and $U_e x - U_{e'} x \in [-\mathbb{1}, \mathbb{1}]$ by (13). Therefore, $\|(e-e')\square x\| \leq 1$. This holds for all $e \in E$. Lemma 9.1 and Lemma 10.1 then imply that $\|y\square x\| \leq 1$ for all $y \in [-\mathbb{1}, \mathbb{1}]$. □

So far it has been seen that the Jordan decomposition property and the absence of third-order interference entail a product $y\square x$ on the order-unit space $A$ generated by the UCP space $E$ which is linked to the conditional probabilities via the identity $U_e x = 2e\square(e\square x) - e\square x$ holding for the events $e$ and the elements $x$ in $A$. This follows from equation (12).

The product $y\square x$ is neither commutative nor associative and thus far away from the products usually considered in mathematical physics. The common product of linear operators is not commutative, but associative and the Jordan product $a \circ b := (ab+ba)/2$ is not associative, but commutative.

## 11 Jordan algebras

The elements of a UCP space $E$ represent the events or propositions; they can be considered also as observables with the simple discrete spectrum $\{0,1\}$ representing a yes/no test experiment. However, what is the meaning of the elements $x$ in the order-unit space $A$ generated by $E$? One might expect that they represent other observables with a larger and possibly non-discrete spectrum and that $\hat{\mu}(x)$ is the expectation value of the observable represented by $x$ in the state $\mu$. If it is assumed that they do so, one would also expect a certain behaviour.

First, one would like to identify the elements $x^2 = x\square x$ and, more generally, $x^n$ (inductively defined by $x^{n+1} := x\square x^n$) in $A$ with the application of the usual polynomial functions $t \to t^2$ or $t \to t^n$ to the observable.

Second, the expectation value of the square $x^2$ should be non-negative; this means that $\hat{\mu}(x^2) \geq 0$ for all $x$ in $A$ and for all states $\mu$, and therefore $x^2 \geq 0$ in the order-unit space $A$.

Third, one would like to have the usual polynomial functional calculus allocating an element $p(x)$ in $A$ to each polynomial function $p$ such that $p_1(x)\square p_2(x) = q(x)$ whenever $p_1$ and $p_2$ are two polynomial functions and the polynomial function $q$ is their product. Since the product in $A$ is not associative, $x^n \square x^m$ need not be identical with $x^{n+m}$. If $x^n \square x^m = x^{n+m}$ holds for all $x$ in $A$ and for all natural numbers $n$ and $m$, $A$ is called power-associative. The availability of the polynomial functional calculus for all elements in $A$ means that $A$ is power-associative, and vice versa.

The following theorem shows that these requirements make $A$ a commutative Jordan algebra; i.e., the product is Abelian and satisfies the Jordan condition $x\square(x^2\square y) = x^2\square(x\square y)$. The Jordan condition is stronger than power-associativity and, in general, power-associativity does not imply the Jordan condition.

**Theorem 11.1:** *Suppose that $E$ is a UCP space with the ε-Hahn-Jordan decomposition property and with $I_3(e_1, e_2, e_3) = 0$ for all orthogonal events $e_1, e_2, e_3$ in $E$. Furthermore, assume that the order-unit space $A$ generated by $E$, together with the multiplication $\square$, is power-associative and that $x^2 \geq 0$ for all $x$ in $A$. Then the multiplication $\square$ is commutative and $A$ is a Jordan algebra. Moreover, $E$ is a $\sigma(A,V)$-dense subset of $\{e \in A : e^2 = e\}$.*

*Proof.* Suppose $x, y \in A$. The positivity of the squares implies the Cauchy-Schwarz inequality $(\hat{\mu}(x \square y))^2 \leq \hat{\mu}(x^2)\hat{\mu}(y^2)$ with states $\mu$. Then, with $y = \mathbb{1}$, $(\hat{\mu}(x))^2 \leq \hat{\mu}(x^2)$. Recall from the proof of Theorem 3.2 that $\|x\| = \sup\{|\hat{\mu}(x)| : \mu \in S\}$. Therefore

$$\|x^2\| \leq \|x\|^2 = sup\{(\hat{\mu}(x))^2 : \mu \in S\} \leq sup\{\hat{\mu}(x^2) : \mu \in S\} = \|x^2\|$$

and hence

$$\|x^2\| = \|x\|^2.$$

Moreover

$$\|x^2 + y^2\| = sup\{\hat{\mu}(x^2) + \hat{\mu}(y^2) : \mu \in S\} \geq sup\{\hat{\mu}(x^2) : \mu \in S\} = \|x^2\|.$$

A second commutative product is now introduced on *A* by $x \circ y := (x \square y + y \square x)/2$. Note that, due to the power-associativity, $x^n$ is identical with the two products. Equipped with this new commutative product, *A* then becomes a Jordan algebra. This follows from a result by Iochum and Loupias [19] (see also [20]).

Since *A* is the dual of *V*, *A* with $\circ$ is a so-called JBW algebra [12]. The JBW algebras represent the Jordan analogue of the W*-algebras and these are the same as the von Neumann algebras, but characterized by a different and more abstract set of axioms. The extreme points of $[0, \mathbb{1}]$ in a JBW algebra are the idempotent elements. Then, by Lemma 10.1, $E \subseteq \{e \in A : e^2 = e\} = ext[0, \mathbb{1}] \subseteq \overline{E}$ and hence *E* is $\sigma(A,V)$-dense in $\{e \in A : e^2 = e\}$.

Suppose $e \in E$. If $\mu$ is any state on *E* with $\mu(e) > 0$, the map $f \to \hat{\mu}(2e \circ (e \circ f) - e \circ f)/\mu(e)$ for $f \in E$ defines a version of the conditional probability (which was shown in [2]) and must coincide with $\hat{\mu}(U_e f)/\mu(e)$. Therefore $U_e f = 2 e \circ (e \circ f) - e \circ f$ and

$$e \square f = T_e f = (f + U_e f - U_{e'} f)/2 = e \circ f$$

for all *e* and *f* in *E*. The product on a JBW algebra is $\sigma(A,V)$-continuous in each component. It then follows in a first step that $e \square y = e \circ y$ for all *e* in *E* and *y* in *A*, and in the second step that $x \square y = x \circ y$ for all *x* and *y* in *A*. Note that the order of the two steps is important because of the asymmetry of the $\sigma(A,V)$-continuity of the product $x \square y$. Thus finally $x \square y = y \square x$. □

When the starting point is the projection lattice *E* in a JBW algebra *M* (e.g., the selfadjoint part of a von Neumann algebra with the Jordan product) without type $I_2$ part, the order-unit space *A* generated by *E* is the second dual $A = M^{**}$ of *M*. It contains *M* by the canonical embedding in its second dual, but is much larger (unless *M* has a finite dimension); *M* is the norm-closed linear hull of *E* in *A*, while *A* is the $\sigma(A,V)$-closed linear hull of *E*.

A rich theory of Jordan algebras is available and most of them can be represented as a Jordan sub-algebra of the self-adjoint linear operators on a Hilbert space. The major exception is the Jordan algebra consisting of the 3×3 matrices with octonionic entries, and the other exceptions are the so-called exceptional Jordan algebras which all relate to this one [12].

A reconstruction of quantum mechanics up to this point has thus been achieved from a few basic principles. The first one is the absence of third-order interference and the second one is the postulate that the elements of the constructed algebra exhibit an behaviour which one would expected from observables. The third one, the ε-Hahn-Jordan decomposition property, is less conceptional and more a technical mathematical requirement.

## *12 Conclusions*

The combination of a simple quantum logical structure with the postulate that unique conditional probabilities exist provides a powerful general theory which includes quantum mechanics as a special case. It is useful for the reconstruction of quantum mechanics from a few basic principles as well as for the identification of typical properties of quantum mechanics that distinguish it from other more general theories. Three such properties have been studied: a novel bound for quantum interference, a symmetry condition for the conditional probabilities, and the absence of third-order interference ($I_3=0$); the third property has been the major focus of this paper.

In the framework of the quantum logics with unique conditional probabilities, the absence of third-order interference ($I_3=0$) has some important consequences. It entails the existence of a product in the order-unit space generated by the quantum logic, which can be used to characterize those quantum logics that can be embedded in the projection lattice in a Jordan algebra. Most of these Jordan algebras can be represented as operator algebras on a Hilbert space, and a reconstruction of quantum mechanics up to this point is thus achieved.

As the identity $I_2=0$ distinguishes the classical probabilities, the identity $I_3=0$ thus characterizes the quantum probabilities. It may be expected that there are other more general theories with $I_3 \neq 0$, and the quantum logics with unique conditional probabilities may provide an opportunity to establish them. For the time being, however, the projection lattices in the exceptional Jordan algebras are the only known concrete examples which do not fit into the quantum mechanical standard model, but still have all the properties discussed in the present paper and do not exhibit third-order interference. Further examples can be expected from Alfsen and Shultz's spectral duality, but unfortunately all the known concrete examples of this theory are either covered by the Jordan algebras or do not satisfy the Gleason-like extension theorem (part (i) of Proposition 3.1). Besides the examples with $I_3 \neq 0$, it would also be interesting to find examples where the identity $I_3=0$ holds, but where the product on the order-unit space is not power-associative or where the squares are not positive or where the symmetry condition (A1) for the conditional probabilities does not hold.

The possibility that no such examples exist is not anticipated, but cannot be ruled out as long as no one has been found. It would mean that every quantum logic with unique conditional probabilities can be embedded in the projection lattice in a Jordan algebra and that third-order interference never occurs. Moreover, the further postulates concerning the behaviour of the observables (power-associativity, positive squares) would then as well become redundant in the reconstruction of quantum mechanics. If this could be proved, the reconstruction process could be cleared up considerably.

In any case, it seems that third-order interference can play a central role in the reconstruction of quantum mechanics from a few basic principles, in an axiomatic access to quantum mechanics with a small number of interpretable axioms, as well as in the characterisation of the projection lattices in von Neumann algebras or their Jordan analogue - the JBW algebras - among the quantum logics. Mathematically, the symmetry property (A1) of the conditional properties (see section 6) can play the same role, which was shown by Alfsen and Shultz [4] and by the author [3]. However, it has a less clear physical meaning than the third-order interference and, therefore, the approach of the present paper seems to be superior form a physical point of view.

**Acknowledgement**. I would like to thank the Perimeter Institute and Philip Goyal for hosting the workshop "Reconstructing Quantum Theory" in August 2009 and Howard Barnum for the inspiring discussions during this workshop and for directing my attention to Sorkin's higher-order interference.